\documentstyle[aps,prl,multicol,epsf]{revtex}

\begin{document}
\draft
\preprint{}

\title{Excess conductance in normal-metal/superconductor junctions}

\author{A. Vaknin, A. Frydman, \thanks{current address: 
Department of Physics, University of California San Diego,
La Jolla, CA 92093} and Z. Ovadyahu} 

\address{The Racah Institute of Physics
The Hebrew University, Jerusalem 91904, Israel}

\maketitle

\begin{abstract}

The current-voltage characteristics of Au/$\text{InO}_{x}$/Pb tunnel junctions exhibit peculiar zero-bias 
anomalies. At low temperatures the zero-bias resistance attains values that are smaller than 
the normal state resistance by a factor that often exceeds 2. The width of this anomaly 
increases with the thickness of the $\text{InO}_{x}$ layer. The possibility that these features arise from 
the nature of the barrier being an Anderson insulator is discussed.  

\end{abstract}


\begin{multicols}{2}

Charge transport through the interface between a normal metal (N) and a superconductor (S) 
is controlled by two processes: Single-particle (Giaever) tunneling, and two-particle 
(Andreev) tunneling \cite{btk}.  Giaever tunneling is the dominant mechanism when the transmission 
coefficient of the NS interface is small and results in current-voltage characteristics such that 
$\text{R}_{0}^{N}/{R}_{0}^{S}$$\ll$1. Andreev tunneling becomes important when the interface 
is ``transparent'' and in 
the limiting case of a ``perfect'' interface may lead to 
$\text{R}_{0}^{N}$/$\text{R}_{0}^{S}$=2. $\text{R}_{0}^{S}$ and $\text{R}_{0}^{N}$ are the interface 
resistance at zero voltage in the superconductor and in the normal state respectively. Both 
types of processes have a characteristic voltage scale of $\Delta$, the superconducting energy gap.
In this note we report on the small-bias I-V characteristics in NIS devices where N is Au, I is 
amorphous indium-oxide, and S is lead. The characteristics of these devices exhibit 
systematic features that cannot be accounted for by current models of Andreev processes in 
NS contacts. We discuss the possibility that these anomalies are peculiar to tunneling through 
localized states of which the barrier (which is an Anderson insulator) is composed.

The Au/$\text{InO}_{x}$/Pb samples were prepared by depositing a gold strip, either 30 $\mu$m or 100 $\mu$m 
wide and 400-500 {\AA} thick onto room-temperature glass-slide. Then, a layer of $\text{InO}_{x}$ 
(thickness L ranging from 90 to 600 {\AA}) was e-beam evaporated on top of the Au electrode. 
Finally a cross strip of Pb 30 $\mu$m or 70 $\mu$m wide and 2000-2500 {\AA} thick completed a standard 
4-terminal device. Fuller details of sample preparation, their structural study, and 
measurements techniques are reported elsewhere \cite{aviad1,aviad2}.

Figure 1 shows the I-V characteristics of a typical NIS sample measured by dc. These are 
compared with the I-V curve expected for an ``ideal'' junction namely, one having a unity 
Andreev coefficient. In the latter case $\text{(dI/dV)}_{S}=\text{2(dI/dV)}_{N}$ 
for $\text{V}\leq\Delta$ and  $\text{(dI/dV)}_{S}=\text{(dI/dV)}_{N}$ for 
$\text{V}>\Delta$. For $\text{V}>\Delta$ $\text{(I-V)}_{S}$ is characterized by a 
constant ``excess-current'' $\text{I}_{S}-\text{I}_{N}$\cite{btk}. S and N 
subscripts are used here to designate measurables in the superconducting and normal state 
respectively. By comparison, the experimental $\text{(I-V)}_{S}$ curve shows 
$\text{(dI/dV)}_{S}\cong2.3\text{(dI/dV)}_{N}$ for 
V=$\pm$0.25 mV and has smaller excess-current at higher bias than the ``ideal''. 

\begin{figure}[b]
\centerline{
\epsfxsize=90mm
\epsfbox{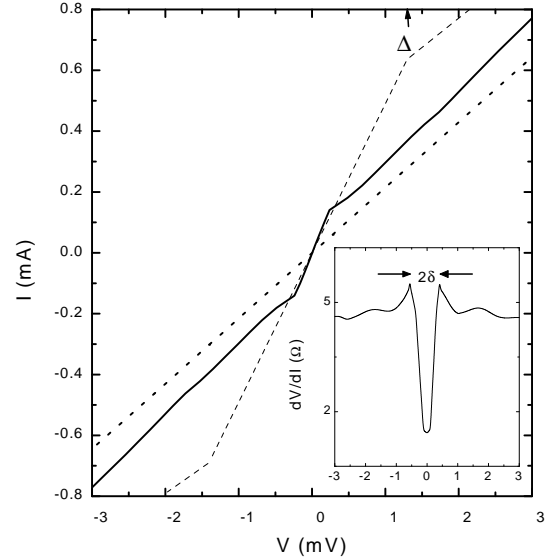}
}
\vspace{-20mm}
\narrowtext
\caption{I-V characteristics of a typical Au/$\text{InO}_{x}$/Pb sample with L=180 {\AA} at the superconducting 
state (full line), and normal state induced by H=0.5 T (dotted line). Data shown were taken 
at T=4.11 K. The dashed line curve depicts the I-V for ``ideal'' Andreev tunneling (see text). 
The inset shows the dynamic resistance of this sample and the definition of $\delta$.}
\end{figure}

The ratio, 
 $\text{R}_{0}^{N}/\text{R}_{0}^{S}$
increases sharply just below the transition temperature $\text{T}_{C}$ of the Pb electrode, and it saturates at 
low temperatures as illustrated in figure 2.  $\text{R}_{0}^{S}$ 
and  $\text{R}_{0}^{N}$  were taken from the respective dV/dI 
plots at zero-bias, always making sure that the excitation current is small enough. The $\text{(dV/dI)}_{N}$ 
plots in the range $\pm$ 20 mV showed only weak structure (few percents in magnitude) that was L 
dependent. For L$\leq$180 {\AA} (L$>250$ {\AA}) a shallow dip (peak) centered at zero-bias was observed. In 
the intermediate regime (i.e., 180 {\AA} $<\text{L}<250$ {\AA}), two shallow dips symmetrical with respect to 
V=0 were often observed with slightly different depths. This non-trivial behavior resembles 
that expected of resonant tunneling through two-level systems discussed by Galperin\cite{galperin}  and by 
Zawadowski\cite{zawadowski} , and could be a relevant for the anomalies we observe.

\begin{figure}[b]
\centerline{
\epsfxsize=90mm
\epsfbox{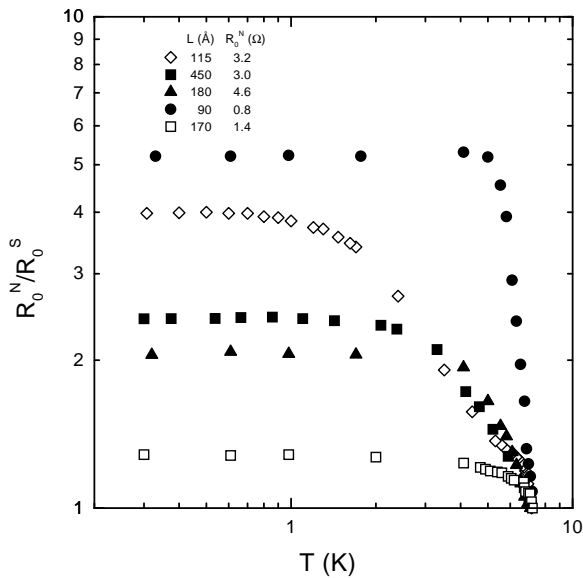}
}
\vspace{-20mm}
\narrowtext
\caption{Temperature dependence of the resistance ratio $\text{R}_{0}^{N}$/$\text{R}_{0}^{S}$ for representative NIS samples.}
\end{figure}

The samples studied in this paper had  $\text{R}_{0}^{N}$ in the range 0.8-5 $\Omega$ at 4K. Between room 
temperature and 100 K  $\text{R}_{0}^{N}$ increased by up to one order of magnitude, depending on L, and 
below 10-20 K  $\text{R}_{0}^{N}$ was temperature independent in all cases. This behavior is consistent with 
transport through a thin Anderson insulator (i.e., the indium-oxide layer) \cite{pollak1}. For T$<7.2$ K, the 
$\text{T}_{C}$ 
of Pb,  $\text{R}_{0}^{N}$ was obtained by applying a magnetic field H parallel to the sample plane to quench 
superconductivity (typically, 0.15 T). At higher fields, up to 3 T,  $\text{R}_{0}^{N}$ 
was essentially unaffected 
by H. Within our experimental error this  $\text{R}_{0}^{N}$ was identical with the zero-bias R measured just 
above $\text{T}_{C}$. There is therefore no reason to believe that the large value of 
 $\text{R}_{0}^{N}/\text{R}_{0}^{S}$ reported here is 
due to overestimating the value of  $\text{R}_{0}^{N}$. The majority of our samples (more than 80 samples 
altogether) had  $\text{R}_{0}^{N}/\text{R}_{0}^{S}$ values between 2 and 4. 
Three samples had  $\text{R}_{0}^{N}/\text{R}_{0}^{S}$ of about 5 and five 
samples had $1<\text{R}_{0}^{N}/\text{R}_{0}^{S}<2$. No correlation was found between 
 $\text{R}_{0}^{N}/\text{R}_{0}^{S}$ and  $\text{R}_{0}^{N}$ or with the 
thickness of the Anderson insulator L. In fact, in a series of 3 samples prepared at the same 
deposition run, and where L was 200 {\AA}, 400 {\AA}, and 600 {\AA}, 
 $\text{R}_{0}^{N}/\text{R}_{0}^{S}$ was essentially the same 
for all three. On the other hand, there seems to be an intriguing correlation between L and the 
{\em range} of voltages, $\delta$, over which the anomalous excess conductance is observed. It turns out that 
the smaller L is the {\em narrower} is this range, a trend that becomes quite apparent for samples with 
L$<200$ {\AA}. Figure 3 compares the dynamic resistance curves vs. bias voltage for a small L 
sample with a larger L sample clearly illustrating this point. A convenient measure of $\delta$ is the 
position of the peak in the dV/dI vs. V plot ({\em c.f}., inset to figure 1). The dependence of $\delta$ on L is 
shown in figure 4. Although there is a considerable scatter in the data the overall trend is clear. 
No such correlation could be identified between $\delta$ and  $\text{R}_{0}^{N}$. 
In particular, samples with identical 
L  but with quite different junction areas (and therefore different  $\text{R}_{0}^{N}$) exhibit similar $\delta$.

\begin{figure}[b]
\centerline{
\epsfxsize=90mm
\epsfbox{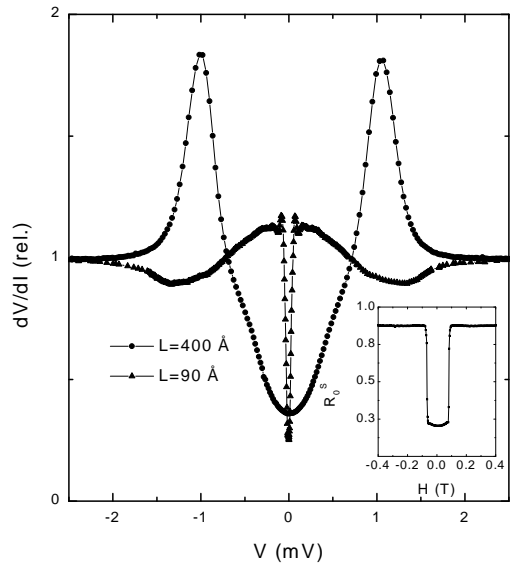}
}
\vspace{-20mm}
\narrowtext
\caption{Dynamic resistance (normalized to the value at 2.5 mV) vs. V for NIS samples with small 
and large L (measured at 4.11K). Note the dips at $\pm\Delta$ for the L=90 {\AA} sample. Inset shows 
the dependence of $\text{R}_{0}^{S}$ on H for the L=90 {\AA} sample.}
\end{figure}

Qualitatively, the zero-bias anomalies we observe have the features expected of NS contacts 
in which Andreev processes are dominant \cite {btk}. In particular, the correlation of the phenomenon 
with the appearance of superconductivity in the Pb electrode ({\em c.f}., figures 2 and 3) is 
suggestive of some sort of a proximity-effect. However, we are not able to account for the 
larger-than-two value of $\text{R}_{0}^{N}$/$\text{R}_{0}^{S}$ or for the dependence of
$\delta$ on L by any of the current 
models for NS contacts. While we cannot rule out structural imperfections, it is hard to see 
how they can explain the observed anomalies. For example, the I-V characteristics are 
incompatible with proximity-induced superconductivity in {\em normal-metal} filaments. The 
resistance of such a sample may go down below $\text{T}_{C}$, but not by more than a factor of two, and 
it should gradually revert to its normal value (rather than saturate) at low temperatures.  Also, 
in N/N'/S samples where N' was a semi-continuous normal metal (a system which should 
closely resemble random array of filaments), the dependence of $\delta$ on L exhibited just the 
opposite trend than observed here \cite{ucsd}.  Finally, the fact that 
$\text{R}_{0}^{S}$ goes to $\text{R}_{0}^{N}$ at the {\em same} H at 
which superconductivity is destroyed in the Pb electrode (figure 3) is inconsistent with 
transport through superconducting filaments \cite{rem1}.
 
\begin{figure}[b]
\centerline{
\epsfxsize=90mm
\epsfbox{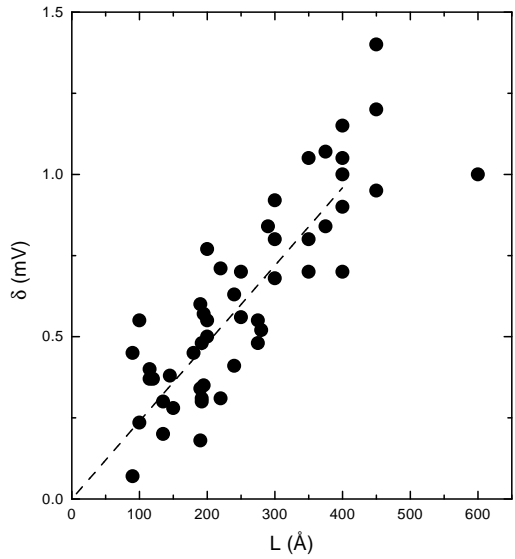}
}
\vspace{-20mm}
\narrowtext
\caption{The value of $\delta$ as function of L for fifty of the studied NIS samples (only samples with 
0.8 $\Omega\leq\text{R}_{0}^{N}\leq5 \Omega$ are included).}
\end{figure}

While we cannot see how ``technological'' defects (that may or may not be there) can account 
for the anomalies we see, there is an inherent feature in the samples that {\em must} be reckoned 
with: The barriers in our samples are Anderson insulators, and that means that they include a 
significant density of localized states. This leads us to consider the possible role of the 
Anderson insulator in this problem. Andreev tunneling through such media received little 
theoretical attention to date so one can assess the relevance of this scenario only in general 
terms. It is established both theoretically \cite{nss} and experimentally \cite{lalko} that coherent quantum 
mechanical effects can be observed in Anderson insulators. At helium temperatures the 
phase-coherent length in such systems is typically \cite{lalko} few hundred {\AA} which is comparable with 
L in our junctions, and therefore coherent tunneling through the medium is feasible. There 
are several indications that suggest that elastic tunneling processes take place in our samples. 
There are several indications that suggest that tunneling 
processes take place in our samples. In dV/dI plots of Pb/$\text{InO}_{x}$/Pb devices with L$<200$ {\AA} we 
observed \cite{aviad2} a dip at V=$\pm2\Delta$ and prominent modulations at V=$\pm$4.5 mV and $\pm$8.5 mV, the 
phonon energies of lead\cite{ady}.  These two features are clear evidence for Giaever tunneling. In the 
NIS devices this tunneling channel manifests itself as a dip in dV/dI at $\pm\Delta$ which can be clearly 
seen in samples with L$<150$ {\AA} ({\em c.f}., figure 3). Finally, {\em all} our samples, 
both NIS and SIS, show 
excess current at least up to voltages that are 5 times $\Delta$ consistent with Andreev tunneling 
processes\cite{btk}.

The correlation between $\delta$ and L may be also related to transport through localized states. 
Note that $\delta\propto$L is consistent with a 
characteristic {\em electric field} $\text{F}_{C}$ at which the process responsible for the large 
 $\text{R}_{0}^{N}/\text{R}_{0}^{S}$ is broken 
or considerably weakened. As 
mentioned above, this feature is not observed in diffusive systems \cite{ucsd}. For Anderson insulators 
however, this result is natural because transport through localized states is very sensitive to 
electric fields \cite{shklovskii}. A typical field associated with 
V=$\delta$ is F$\cong$250 V/cm (estimated from the straight line in figure 4). Fields of these magnitudes are 
sufficient to appreciably reduce quantum-coherent effects associated with ``forward-scattered'' 
tunneling paths in Anderson insulators\cite{faran}.  Andreev processes are even more sensitive 
to electric fields than single-particle 
processes because the electron and hole acquire different phases upon traversing it. This may 
become important in multiple scattering situations: As pointed out by van-Wees et al \cite{van},  the 
Andreev coefficient at the NS interface is enhanced by disorder due to a constructive 
interference between the electron and hole. 
This enhancement effect persists up to a critical voltage 
$\text{V}_{C}$ 
given by$\hbar\text{/e}\tau_{\varphi}$ ($\tau_{\varphi}$ is the phase-breaking time of the disordered 
region at which the particle is 
``trapped'' near the interface).
For NS contacts this should lead to a 
zero-bias anomaly with voltage-width that {\em decreases} with L.

The situation in the NIS case, however, is different than that of the NS system in an essential 
way. To see that, consider a semi-infinite S and N layers separated by a layer of an Anderson 
insulator I of extent L along the Z-axis such that the SI interface is at Z=0 and the NI 
interface at Z=L. The electronic states in I are localized on scale $\xi$ much smaller than L. This 
introduces a natural hierarchy in the problem. States with Z$\leq$0 are strongly coupled to S 
because on such scales there is no distinction between an Anderson insulator and a (dirty) 
normal metal. The conditions for the van-Wees et al mechanism are obeyed for this thin layer 
and therefore pairing-amplitude will be induced in it \cite{rem2}.  On the other hand, the states in N 
(that are responsible for the measured tunneling transport) have an exponentially small 
coupling to the superconductor because L$\gg\xi$. But this exponential coupling also makes the 
N$\rightarrow$S tunneling sensitive to modifications in the nature of the intermediate states in I, 
including those in $0<\text{Z}\leq\xi$. Now, the modification in the interface layer (and its effect on the 
I-V characteristics) is cut-off when the voltage drop {\em across} it exceeds $\text{V}_{C}$. In other words we 
associate $\delta$ with the voltage across L that imposes a field $\text{F}_{C}={V}_{C}$$\xi$. 
This immediately leads to 
the dependence $\delta\propto$L. To check on the plausibility of this approach note that using $\xi$=10 {\AA} as 
a typical value for the localization length in our samples \cite{aviad2}, and 
$\text{F}_{C}$=250 V/cm gives 
$\tau_{\varphi}$=$\hbar$/(e$\text{F}_{C}\xi\text{)}\approx$2$\cdot$${10}^{-11}$ seconds which is a reasonable value for the inelastic time at these 
temperatures and fields \cite{rem3}.

Following van-Wees et al \cite{van}, it may be argued that applying a magnetic field should destroy 
the interference and thus weaken the zero-bias anomaly. But, the smallness of $\xi$ makes this 
field larger than that necessary to quench superconductivity in the Pb electrode. An ac field 
of frequency comparable to $\tau_{\varphi}^{-1}$ may be a more effective de-phasing agent in this case. We 
have indeed observed a dramatic reduction of ($\text{R}_{0}^{S})^{-1}$ in samples exposed to a 20 GHz 
microwave source. This reduction in the zero-bias conductance was much too big to be 
explained by the barely measurable Joule-heating of the sample due to the microwave field \cite{rem4}.  
Full details of the microwave experiments will be given elsewhere \cite{micro}.
 
This heuristic picture offers then a plausible way to understand the origin of the dependence 
of $\delta$ on L. The key question that remains to be answered is whether it can also account for the 
observation of $\text{R}_{0}^{N}$/$\text{R}_{0}^{S}>2$. This must await a detailed theoretical treatment of these issues. In 
particular the nature of the proximity-effect in Anderson insulators needs to be better 
understood. We hope that the present results will motivate such studies.
We gratefully acknowledge useful discussions with M. Pollak. This research was supported 
by a grant administered by the Israel Science Foundation.

\end{multicols}  

\begin{references}
\bibitem{rem1} A superconducting filament, whether extending across all or part of the junction width ought to 
have a much larger $\text{H}_{C}$ than that of the thick Pb electrode.
\bibitem{rem3} At T$\leq$4 K and F$\geq$100 V/cm the hopping length (and thus $\tau_{\varphi}$) is controlled 
by F rather than by T (see references 14)
\bibitem{rem4} The shift in the sample temperature can be accurately measured by the position of the 
`above-gap' structure which scales with $\Delta$(T), see reference 12 for details.
\end{references}
\end{document}